\documentclass[%
twocolumn,
 amsmath,amssymb,
 aps,
 showpacs
]{revtex4-1}

\usepackage{amsmath}
\usepackage{mathrsfs}
\usepackage{graphicx}
\usepackage{dcolumn}
\usepackage{bm}
\usepackage{ulem}               

\begin{document}

\preprint{APS/123-QED}

\title{
Time-dependent restricted-active-space self-consistent-field theory 
for laser-driven many-electron dynamics
}

\author{Haruhide Miyagi}
\author{Lars Bojer Madsen}
\affiliation{Department of Physics and Astronomy, Aarhus University, 8000 {\AA}rhus C, Denmark }

\date{\today}

\begin{abstract}
We present the time-dependent restricted-active-space self-consistent-field (TD-RASSCF) theory as a new framework for the time-dependent many-electron problem. The theory generalizes the multiconfigurational time-dependent Hartree-Fock (MCTDHF) theory by incorporating the restricted-active-space scheme well known in time-independent quantum chemistry. 
Optimization of the orbitals as well as the expansion coefficients at each time step makes it possible to construct the wave function accurately while using only a relatively small number of electronic configurations. In numerical calculations of high-order harmonic generation spectra of a one-dimensional model of atomic beryllium interacting with a strong laser pulse, the TD-RASSCF method is reasonably accurate while largely reducing the computational complexity. The TD-RASSCF method has the potential to treat large atoms and molecules beyond the capability of the MCTDHF method.
\end{abstract}

\maketitle


\section{Introduction}

Development of reliable approximate theories for the description of time-dependent many-electron dynamics has been desirable for decades, and its importance is especially emphasized nowadays by the need of support to experiments on the real time analysis and control of ultrafast electronic and nuclear dynamics of atoms and molecules by intense laser pulses \cite{Baker2006, Pfeiffer2012, Shafir2012, Blaga2012, Krausz2009}. In numerical simulations, however, the problems are most often simplified by the single-active-electron (SAE) approximation \cite{Kulander1992}, which assumes that only one electron is moving in an effective potential. In theoretical approaches, the combination of the SAE and the strong field approximations, where the interaction with the atomic or molecular potential is treated perturbatively, has been widely accepted as a standard approach in this research area. The Lewenstein model \cite{Lewenstein1994}, which is built on these assumptions, makes it possible to easily compute high-order harmonic generation (HHG) spectra of atoms and molecules and also provides a clear physical picture based on the semiclassical three-step model \cite{Krause1992,Schafer1993,Corkum1993}. While the studies within the framework of the SAE approximation have succeeded in providing a qualitative understanding of phenomena, multielectron effects are also recognized to play a crucial role, e.g., in time delay studies of photoionization \cite{Schultze2010,Klunder2011}, and moreover multiple orbital contributions to HHG spectra are widely observed for atoms and molecules \cite{Jordan2008,Smirnova2009,Sukiasyan2009,Sukiasyan2010}. To describe many-electron dynamics, several \textit{ab initio} approaches beyond the SAE approximation have been developed. Among others, the time-dependent $R$-matrix method is one of the most elaborate ways for describing single electron ionization processes and taking into account the electron correlation \cite{Lysaght2009, Moore2011, Brown2012}. One of the computationally and conceptually simpler approaches is the time-dependent configuration-interaction singles (TD-CIS) method \cite{Rohringer2006, Gordon2006, Rohringer2009, Greenman2010, Pabst2012}, in which the CI-expansion is truncated at singly excited configurations relative to the Hartree-Fock ground state. Both these methods can be considered to be special cases of a more generalized concept, namely, the time-dependent restricted-active-space configuration-interaction (TD-RASCI) method \cite{Hochstuhl2012}. 

Over the last decade, originating from the time-dependent Hartree-Fock theory \cite{Kulander1987,Pindzola1995,Dahlen2001}, a more sophisticate framework called multiconfigurational time-dependent Hartree-Fock (MCTDHF) theory has been developed and quite recently shown its potential for analyzing ultrafast laser driven electron dynamics in atoms and molecules \cite{Jordan2008,Sukiasyan2009,Sukiasyan2010,Caillat2005, Nest2007, Nest2008, Jordan2008, Kato2008, Kato2009a, Hochstuhl2011, Haxto2012}, and moreover for elucidating the role of electron-nuclear correlation in a molecule during ionization \cite{Kato2009, Nest2009, Jhala2010,Haxto2011, Ulusoy2012} (see also references on the multiconfigurational time-dependent Hartree (MCTDH) theory, e.g., the original paper \cite{Meyer1990}, a review \cite{Beck2000}, and a textbook \cite{Meyer2010}. In addition, multiconfigurational theory has been explored for bosonic systems \cite{lenz}). In the MCTDHF theory with the spin restricted ansatz, the $N_{\rm e}$-electron wave function is expressed by
\begin{eqnarray}
|\Psi(t)\rangle
=
\sum_{I\in\mathcal{V}_{\rm FCI}}C_I(t)|\Phi_I(t)\rangle,
\label{Rwave}
\end{eqnarray}
where $|\Phi_I(t)\rangle$ denotes a normalized Slater determinant for $N_{\rm e}$ electrons built from a set of time-dependent active spin-orbitals $\{\phi_i(t)\}_{i=1}^{2M}$, and the multi-index $I=(i_1,\cdots,i_{N_{\rm e}})$ is a string of orbital indices of which the Fock space is composed: $\mathcal{V}_{\rm FCI}=\big\{(i_1,\cdots,i_{N_{\rm e}})|1\le i_1<\cdots<i_{N_{\rm e}}\le 2M\big\}$. Using the Dirac-Frenkel-McLachlan time-dependent variational principle \cite{Dirac1930, Frenkel1934, McLachlan1964, Lubich2008}, the equations of motion are derived which optimize the orbitals as well as the expansion coefficients in each time step. This optimization procedure leads to the expectation that the system can be accurately described with a relatively small number of orbitals, $2M$. However, because the Fock space $\mathcal{V}_{\rm FCI}$ in the MCTDHF theory is spanned by all possible configurations for a given set of spin-orbitals, the computational cost due to the expansion coefficients $\{C_I\}_{I\in\mathcal{V}_{\rm FCI}}$ is proportional to the number of ways that $N_{\rm e}$ electrons can be distributed in the $2M$ spin-orbitals
\begin{eqnarray}
\dim\big(\mathcal{V}_{\rm FCI}\big)
=
\left(
\begin{array}{c}
M\\
N_{\rm e}/2 \\
\end{array}
\right)^2
\alt
O\big(M^{N_{\rm e}}\big),
\label{total_wave_CI}
\end{eqnarray}
i.e., roughly speaking, exponentially scaling with respect to the number of electrons, $N_{\rm e}$. Hence for the investigation of the nonperturbative laser driven electron dynamics, this unfavorable scaling with $N_{\rm e}$ impedes the MCTDHF method to be extended to systems with more than a few electrons, i.e., beyond systems like He \cite{Hochstuhl2011}, Be \cite{Haxto2012}, H$_2$ \cite{Kato2008, Kato2009a, Haxto2011}, and LiH \cite{Nest2007, Nest2008,Ulusoy2012}. In order to carry on the study for larger systems, it is therefore inevitable to abandon the full CI-expansion of Eq.~\eqref{Rwave}.

For the time-dependent problem, it is natural to follow standard approximations in quantum chemistry such as, for example, CIS and explore their time-dependent counterparts, the TD-CIS method \cite{Rohringer2006, Gordon2006, Rohringer2009, Greenman2010, Pabst2012}. It is thus natural as well to investigate the possibility of a truncation of the expansion in the MCTDHF theory at a specific excitation level. Within the framework of the MCTDH theory, the truncation has already been explored \cite{Worth2000}. Our study is hence aiming at a further generalization of the MCTDHF theory by incorporating it with the RAS approach; decomposing the single-particle Hilbert space into several subspaces, among which electron transitions are allowed with several restrictions. The new framework is hereafter referred to as the time-dependent restricted-active-space self-consistent-field (TD-RASSCF) theory. It is emphasized that key ingredients of the theory are (i) use of time-dependent orbitals and (ii) truncation of the CI-expansion. Despite of its simple concept, the truncation of the expansion partly destroys the principal bundle structure inherent in the MCTDHF theory \cite{Lubich2008,Kvaal2011}. Accomplishing the truncation hence requires a careful analysis of the structure of the theory. A consistent formulation of the TD-RASSCF theory is the main purpose of the present work. 

This study is inspired by a recent work \cite{Kvaal2012}, where another new method called orbital adaptive time-dependent coupled-cluster (OATDCC) theory was formulated. In this method, the CC-expansion is truncated at doubly excited configurations, but higher excitations are also partly included due to the nonlinear property inherent in the CC ansatz (see, e.g., Refs.~\cite{Helgaker2000,Komarov2009} for a discussion of time-independent CC theory). Furthermore, the CC-expansion ensures the size-consistency and -extensivity, which will be of utmost importance for correctly describing dissociation processes of molecules. However, there are still some problems remaining: because the left- and right-wave functions in the OATDCC theory are parametrized in different ways, imaginary time relaxation is not readily feasible as a means to calculate the ground state wave function needed for the following real time analysis of the dynamics. It is thus attractive to develop methods not suffering from these complications while only slightly compromising the accuracy; the TD-RASSCF theory is one such example.

The paper is organized as follows. The TD-RASSCF theory is formulated in Sec.~\ref{GFormulation}. The working equations are derived based on the time-dependent variational principle combined with the Lagrange multiplier method. Explicit forms of the equations are given and compared to the corresponding ones in the MCTDHF theory. A central aspect of the formulation is given in Sec.~\ref{Pspace}, where we concentrate on a discussion of the parametric redundancy in the time-dependent SCF theory. As a proof-of-principal application of the theory, one-dimensional (1D) model atoms are investigated in Sec.~\ref{Numerical tests}: calculations of the ground state wave function, followed by computations of the HHG spectra. The analysis of the convergence property of the TD-RASSCF calculations also uncovers the time-dependent many-electron dynamics. Section \ref{Conclusion} provides a summary and concludes. A discussion of orbital rotations and the accompanying simplifications in the case of a two-electron system is deferred to Appendix A.

\section{\label{GFormulation} Formulation}

\begin{figure}
\begin{center}
\begin{tabular}{c}
\resizebox{80mm}{!}{\includegraphics{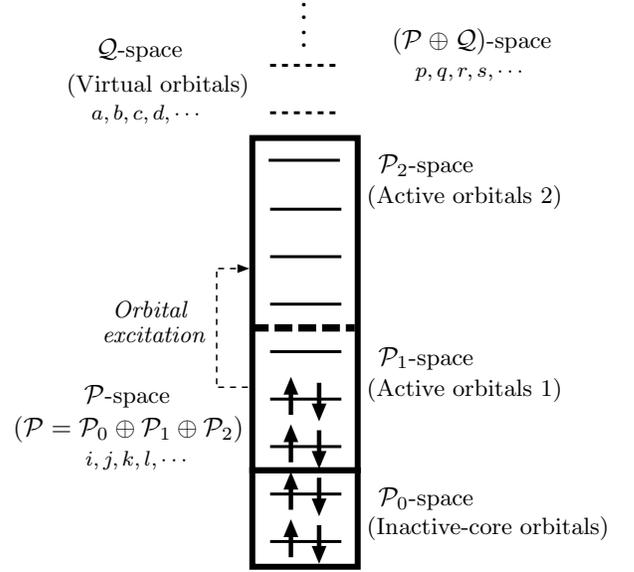}}
\end{tabular}
\caption{
\label{fig_orbital_ras}
Illustration of the division of the single-particle Hilbert space in the TD-RASSCF theory. The wave function is expand by using orbitals in $\mathcal{P}$-space, which is composed by an inactive-core space, $\mathcal{P}_0$, and two active spaces, $\mathcal{P}_1$ and $\mathcal{P}_2$, where a partition exists through which electrons can transit with several restrictions. In accordance with the convention in the MCTDHF theory, the rest of the single-particle Hilbert space spanned by the virtual orbitals is referred to as $\mathcal{Q}$-space. The orbitals in either $\mathcal{P}$- or $\mathcal{Q}$-space are labeled by $p,q,r,s,\cdots$, while the $\mathcal{P}$-space orbitals are labeled by $i,j,k,l,\cdots$, and the $\mathcal{Q}$-space orbitals by $a,b,c,d,\cdots$. 
}
\end{center}
\end{figure}

Consider an $N_{\rm e}$-electron system described by a generic time-dependent Hamiltonian including one- and two-body operators. In second quantization, the Hamiltonian reads
\begin{eqnarray}
H(t)
=
\sum_{pq} h^p_q(t) c_p^{\dagger}c_q
+
\frac{1}{2}
\sum_{pqrs} v^{pr}_{qs}(t) c_p^{\dagger}c_r^{\dagger}c_sc_q,
\label{Hamiltonian1}
\end{eqnarray}
where $c_p$ ($c_p^{\dagger}$) is the annihilation (creation) operator of an electron in the spin-orbital $|\phi_p(t)\rangle$, satisfying the anticommutation relation $\{c_p,c_q^{\dagger}\}=\delta^p_q$. The prefactors of the operators in the Hamiltonian read in first quantization:
\begin{eqnarray}
h^{p}_{q}(t)&=&\int\phi_p^{\dagger}(z,t)h({\bm r},t)\phi_q(z,t) dz,
\label{one-body1}
\end{eqnarray}
and
\begin{eqnarray}
v^{pr}_{qs}(t)=\iint&&\phi_p^{\dagger}(z_1,t)\phi_r^{\dagger}(z_2,t)
\nonumber \\
&&\times
v({\bm r}_1,{\bm r}_2)\phi_q(z_1,t)\phi_s(z_2,t) dz_1dz_2,
\label{two-body1}
\end{eqnarray}
where the spin-orbitals are represented in the spin and spatial coordinates $z=({\bm r},\sigma)$. The two-body operator $v({\bm r}_1,{\bm r}_2)$ denotes the Coulomb repulsion between two electrons, and the one-body operator $h({\bm r},t)$ depends on time via dipole interactions with external laser fields. In the TD-RASSCF theory, the $N_{\rm e}$-electron wave function is expressed by
\begin{eqnarray}
|\Psi(t)\rangle
=
\sum_{I\in\mathcal{V}_{\rm RAS}}C_I(t)|\Phi_I(t)\rangle
\label{MCTDHFK_wave_function}
\end{eqnarray}
in which, differently from Eq.~(\ref{Rwave}), the Fock space $\mathcal{V}_{\rm RAS}$ is now composed of several selected configurations and not the full configuration space. In this study, the orbitals are classified as in Fig.~\ref{fig_orbital_ras}: the single-particle Hilbert space is divided into two subspaces, $\mathcal{P}$-space contributing to the construction of the wave function, i.e., the multi-index $I$ in Eq.~\eqref{MCTDHFK_wave_function} contains 
indices of $\mathcal{P}$-space orbitals, and the supplementary virtual orbital space hereafter called $\mathcal{Q}$-space. The indices $p,q,r,s\cdots$ denote orbitals belonging to either space, while the $\mathcal{P}$-space orbitals are labeled by $i,j,k,l,\cdots$, and the virtual $\mathcal{Q}$-space orbitals by $a,b,c,d,\cdots$. The RAS scheme is based on dividing the $\mathcal{P}$-space into three subspaces usually denoted by RAS1, 2, and 3. In the original and the most general definition, the RAS1 and 3 spaces are characterized by the minimal and maximal occupation numbers, respectively, and the RAS2 space has no constraint \cite{Olsen1988,Malmqvist1990,Helgaker2000}. The RAS scheme in this paper is, however, supposing more specific cases as shown in Fig.~\ref{fig_orbital_ras}. Here the $\mathcal{P}$-space is composed of an inactive-core space, $\mathcal{P}_0$, and two active spaces, $\mathcal{P}_1$ and $\mathcal{P}_2$, between which electron transitions are allowed with several restrictions (see Sec.~IV).

For describing the dynamics within this framework, we need the set of equations obeyed by the expansion coefficients and the orbitals. To this end, we follow a standard prescription and use the Dirac-Frenkel-McLachlan time-dependent variational principle \cite{Dirac1930, Frenkel1934, McLachlan1964, Lubich2008}. Henceforth the explicit time-dependence of the parameters and the operators is dropped for brevity as long as this ease of notation will not lead to confusion. First we define an action functional (atomic units are used throughout)
\begin{eqnarray}
\mathcal{S}&&\big[\{C_I\},\{\phi_i\},\{\varepsilon^i_j\}\big]=
\nonumber \\
&&\int_0^T
\Bigg[
\langle\Psi|\Bigg( i\frac{\partial}{\partial t}-H\Bigg)
|\Psi\rangle
+\sum_{ij} \varepsilon^i_j(t)\Big(
\langle\phi_i|\phi_j\rangle
-\delta^i_j
\Big)
\Bigg]dt.
\nonumber \\
\label{functional_CI}
\end{eqnarray}
Then we use that a stationary point 
\begin{eqnarray}
\delta\mathcal{S}\big[\{C_I\},\{\phi_i\},\{\varepsilon^i_j\}\big]=0
\end{eqnarray}
provides the best approximation of the dynamics for the given ansatz. Here $\varepsilon^i_j$ is the Lagrange multiplier that ensures orthonormality of the $\mathcal{P}$-space orbitals during the time interval $[0,T]$. The variation of the action functional is
\begin{eqnarray}
\delta
\mathcal{S}&&\big[\{C_I\},\{\phi_i\},\{\varepsilon^i_j\}\big]
=
\nonumber \\
&&
\int_0^T
\Bigg[
\langle\delta\Psi|\Bigg( i\frac{\partial}{\partial t}-H\Bigg) |\Psi\rangle
+
\langle\Psi|\Bigg( -i\overleftarrow{\frac{\partial}{\partial t}}-H\Bigg)|\delta\Psi\rangle
\nonumber \\
&&
\hspace{20mm}
+\sum_{ij} \varepsilon^i_j\Big(
\langle\delta\phi_i|\phi_j\rangle
+\langle\phi_i|\delta\phi_j\rangle
\Big)
\nonumber \\
&&
\hspace{20mm}
+\sum_{ij} \delta\varepsilon^i_j\Big(
\langle\phi_i|\phi_j\rangle
-\delta^i_j
\Big)
\Bigg]dt,
\label{varitaion1_2}
\end{eqnarray}
where, performing integration by parts, a time-derivative operator with a leftward-arrow is introduced to denote its action on the bra-vector. The variation of the wave function (\ref{MCTDHFK_wave_function}) is written as 
\begin{eqnarray}
|\delta\Psi\rangle
=
\sum_{I\in\mathcal{V}_{\rm RAS}} \delta C_I |\Phi_I\rangle
+
\sum_{pq}
c^{\dagger}_p c_q
|\Psi\rangle
\langle \phi_p|\delta\phi_q\rangle.
\end{eqnarray}

Firstly, imposing $\delta\mathcal{S}/\delta\varepsilon^i_j=0$ leads to
\begin{eqnarray}
\langle\phi_i|\phi_j\rangle
=\delta^i_j, 
\label{Lagrange_multiplier}
\end{eqnarray}
which ensures the orthonormality of the $\mathcal{P}$-space orbitals at all times. Stationary conditions with respect to small variations of the other parameters
\begin{eqnarray}
\delta\mathcal{S}/\delta C_I^*=
\delta\mathcal{S}/\langle\delta\phi_i|=0
\label{Stationary_condition1}
\end{eqnarray}
give us the equations of motion. Since the left- and right-wave functions are hermitian conjugates of each other, the stationary conditions $\delta\mathcal{S}/\delta C_I=\delta\mathcal{S}/|\delta\phi_i\rangle=0$ result in a set of equations which is the hermitian conjugate of the set obtained from Eq.~(\ref{Stationary_condition1}).

\subsection{Derivation of the amplitude equations}

The stationary condition $\delta\mathcal{S}/\delta C_I^*=0$ in Eq.~(\ref{Stationary_condition1}) results in a formal expression of the amplitude equations
\begin{eqnarray}
\langle\Phi_I |\left( i\frac{\partial}{\partial t}-H\right)|\Psi\rangle=0.
\label{amp_eqs0}
\end{eqnarray}
Here, the time-derivative of the right-wave function is decomposed into two parts:
\begin{eqnarray}
\frac{\partial}{\partial t}|\Psi\rangle
=
\sum_{I\in\mathcal{V}_{\rm RAS}}
\dot{C}_I|\Phi_I\rangle
+
D|\Psi\rangle,
\label{right_D}
\end{eqnarray}
with
\begin{eqnarray}
D
=\sum_{pq}\eta_q^p c_p^{\dagger}c_q,
\label{definition_D}
\end{eqnarray}
and $\eta^p_q=\langle\phi_p|\dot{\phi}_q\rangle$. This anti-hermitian matrix $\eta^p_q$ plays an important role in the formulation of the orbital equations as we discuss in Sec.~\ref{Pspace}. We insert Eq.~(\ref{right_D}) into Eq.~(\ref{amp_eqs0}) and obtain
\begin{eqnarray}
i\dot{C}_I
+\langle\Phi_I|(iD-H)|\Psi\rangle=0.
\label{amp_eqs1}
\end{eqnarray}
Furthermore, we substitute Eqs.~(\ref{Hamiltonian1}) and (\ref{definition_D}) into Eq.~(\ref{amp_eqs1}) and derive after some algebra the explicit form
\begin{eqnarray}
i\dot{C}_I
&=&
\sum_{ij}{\rm sgn}(\tau)C_{\tau(I_i^j)} (h^i_j -i\eta_j^i) 
\nonumber \\
&&+
\frac{1}{2}
\sum_{ijkl} {\rm sgn}(\tau) C_{\tau(I_{ik}^{jl})} v^{ik}_{jl},
\label{amp_orb}
\end{eqnarray}
with $C_{I_i^j}=\langle\Phi_I|c_i^{\dagger}c_j|\Psi\rangle$, $C_{I_{ik}^{jl}}=\langle\Phi_I|c_i^{\dagger}c_k^{\dagger}c_lc_j|\Psi\rangle$, and $\tau$ a permutation map rearranging strings of orbital indices to ascending order with the sign defined by ${\rm sgn}(\tau)=1$ $(-1)$ when $\tau$ is even (odd). The amplitude equations of Eq.~(\ref{amp_orb}) are exactly the same as those of the MCTDHF theory. In the MCTDHF theory, one can choose all of the $\eta_j^i$ to zero and thereby make the amplitude equations easier to solve (see, e.g., Ref.~\cite{Caillat2005}). As will be shown in Sec.~\ref{Pspace}, in the TD-RASSCF theory all the $\eta_j^i$ can not be set to zero due to the truncation of the CI-expansion, and one thus needs a more careful simultaneous optimization of the expansion coefficients and the orbitals.

\subsection{Derivation of the orbital equations}

The other stationary condition $\delta\mathcal{S}/\langle\delta\phi_i|=0$ in Eq.~(\ref{Stationary_condition1}) using Eq.~(\ref{right_D}) yields the set of equations to be solved for the orbitals
\begin{eqnarray}
\sum_q
|\phi_q\rangle
\langle\Psi_i^q|
\Bigg(
&&
i\sum_{I\in\mathcal{V}_{\rm RAS}}\dot{C}_I|\Phi_I\rangle
+(iD-H)|\Psi\rangle
\Bigg)
\nonumber \\
&&+\sum_{j} |\phi_j\rangle \varepsilon_j^i=0,
\label{orb1}
\end{eqnarray}
where the \textit{one-particle-one-hole state} $\langle\Psi_i^q|\equiv\langle\Psi|c_i^{\dagger}c_q$ is introduced. The orbital equations need to be solved in both the $\mathcal{P}$- and $\mathcal{Q}$-spaces as indicated by the use of the index $q$. Defining projection operators onto the $\mathcal{P}$- and $\mathcal{Q}$-spaces by
\begin{eqnarray}
P&=&
\sum_{i=1}^{2M} |\phi_i\rangle\langle\phi_i|,
\\
Q&=&1-P,
\end{eqnarray}
respectively, the time derivative of each orbital is decomposed into two parts, i.e., contributions from the $\mathcal{P}$- and $\mathcal{Q}$-spaces,
\begin{eqnarray}
|\dot{\phi}_i\rangle
&=&
(P+Q)|\dot{\phi}_i\rangle \nonumber \\
&=&
\sum_{j=1}^{2M} |\phi_j\rangle \eta^j_i
+Q|\dot{\phi}_i\rangle.
\label{total_0}
\end{eqnarray}
It is clearly seen from Eq.~(\ref{total_0}) that the optimization of the active orbitals makes the $\mathcal{P}$-space, and thereby the $\mathcal{Q}$-space, time-dependent. By allowing the orbitals to be time-dependent a relatively small number of active orbitals is sufficient for the accurate expansion of the wave function. When the system is irradiated by a laser pulse, as discussed in Sec.~\ref{HHG}, the $\mathcal{P}$-space ensures the inclusion of the most important part of the continuum as well as bound states for the description of ionization

\subsubsection{$\mathcal{Q}$-space orbital equations}

One can obtain a formal expression of the $\mathcal{Q}$-space orbital equations by multiplying Eq.~(\ref{orb1}) by a virtual orbital bra-vector $\langle\phi_a|$ from the left, and by using the orthogonality between the active and virtual orbitals, 
\begin{eqnarray}
\langle\Psi_i^a|(iD-H)|\Psi\rangle
=0.
\label{Q_orb_eqs0}
\end{eqnarray}
Equation (\ref{Q_orb_eqs0}) is a generalization of Brillouin's theorem \cite{Helgaker2000} to time-dependent problems. Substituting Eqs.~(\ref{Hamiltonian1}) and (\ref{definition_D}) into Eq.~(\ref{Q_orb_eqs0}) and perfoming some algebra with the help of Wick's theorem \cite{Komarov2009}, 
\begin{eqnarray}
c_i^{\dagger}c_a c_p^{\dagger}c_q
&=&
c_i^{\dagger}c_p^{\dagger}c_qc_a+\delta_a^p c_i^{\dagger}c_q,
\\
c_i^{\dagger}c_a c_p^{\dagger}c_r^{\dagger} c_sc_q
&=&
c_i^{\dagger}c_p^{\dagger}c_r^{\dagger} c_sc_q c_a
+\delta_a^p c_i^{\dagger}c_r^{\dagger}c_sc_q
-\delta_a^r c_i^{\dagger}c_p^{\dagger}c_sc_q,
\nonumber \\
\label{Wick1}
\end{eqnarray}
the $\mathcal{Q}$-space orbital equations read
\begin{eqnarray}
\sum_j  (i\eta^a_j -h^a_j) \rho^j_i
=
\sum_{jkl} v^{ak}_{jl}\rho^{jl}_{ik},
\label{Q_orbital_eq0}
\end{eqnarray}
with the density matrices $\rho^j_i$ and $\rho^{jl}_{ik}$ defined by
\begin{eqnarray}
&&\rho^j_i
=
\langle\Psi|c^{\dagger}_ic_j|\Psi\rangle
=
\sum_{I\in\mathcal{V}_{\rm RAS}} {\rm sgn}(\tau)
C^*_{\tau(I_j^i)}C_I,
\label{density2}
\end{eqnarray}
and
\begin{eqnarray}
&&\rho^{jl}_{ik}
=
\langle\Psi|c^{\dagger}_ic^{\dagger}_k c_lc_j|\Psi\rangle
=
\sum_{I\in\mathcal{V}_{\rm RAS}}{\rm sgn}(\tau) C^*_{\tau(I^{ik}_{jl})}C_I.
\label{density3}
\end{eqnarray}
To circumvent explicit numerical treatments of the virtual orbitals, we use the $\mathcal{Q}$-space projection operator and express Eq.~(\ref{Q_orbital_eq0}) as
\begin{eqnarray}
i\sum_j Q|\dot{\phi}_j\rangle\rho^j_i
=\sum_jQh(t)|\phi_j\rangle\rho^j_i
+\sum_{jkl}QW^k_l|\phi_j\rangle
\rho^{jl}_{ik},
\nonumber \\
\label{Q_orbital_eq1}
\end{eqnarray}
where the mean-field operator is defined by
\begin{eqnarray}
W^k_l({\bm r})=\int\phi_k^{\dagger}(z')v({\bm r},{\bm r}')\phi_l(z') dz'.
\end{eqnarray}
We now arrive at formally the same $\mathcal{Q}$-space orbital equations as in the MCTDHF theory (see, e.g., Eq.~(12) in Ref.~\cite{Hochstuhl2011}). The density matrices in Eqs.~(\ref{density2}) and (\ref{density3}) are, however, now calculated based on the RAS scheme.

\subsubsection{$\mathcal{P}$-space orbital equations}

We obtain a set of equations for the $\mathcal{P}$-space orbitals when we multiply Eq.~(\ref{orb1}) by an active orbital bra-vector $\langle\phi_j|$ from the left, 
\begin{eqnarray}
\langle\Psi^j_i|(iD-H)|\Psi\rangle
+
i\sum_{I\in\mathcal{V}_{\rm RAS}}
\langle\Psi^j_i|\Phi_I\rangle \dot{C}_I
+\varepsilon_j^i=0.
\label{orb_eqs_A_right1}
\end{eqnarray}
Equation (\ref{orb_eqs_A_right1}), however, still contains a Lagrange multiplier $\varepsilon_j^i$. Similarly, from the stationary condition $\delta\mathcal{S}/|\delta\phi_j\rangle=0$, or by taking the hermitian conjugate of Eq.~(\ref{orb_eqs_A_right1}) followed by exchanging $i$ and $j$, we arrive at equations containing the same multiplier
\begin{eqnarray}
\langle\Psi|(iD-H)|\Psi_j^i\rangle
-
i\sum_{I\in\mathcal{V}_{\rm RAS}}
\dot{C}^*_I
\langle\Phi_I|\Psi^i_j\rangle
+\varepsilon_j^i=0.
\label{orb_eqs_A_left1}
\end{eqnarray}
Subtraction of Eq.~(\ref{orb_eqs_A_right1}) from Eq.~(\ref{orb_eqs_A_left1}) removes the multiplier and gives the $\mathcal{P}$-space orbital equations or the time-dependent Brillouin's theorem for the active orbitals
\begin{eqnarray}
\langle\Psi|(iD-H)|\Psi_j^i\rangle
-
\langle\Psi^j_i|(iD-H)|\Psi\rangle
=
i\dot{\rho}^j_i,
\label{orb_eqs_P}
\end{eqnarray}
where the time-derivative of the density matrix is
\begin{eqnarray}
\dot{\rho}^j_i=
\sum_{I\in\mathcal{V}_{\rm RAS}}
\left(
\dot{C}^*_I
\langle\Phi_I|\Psi^i_j\rangle
+
\langle\Psi^j_i|\Phi_I\rangle\dot{C}_I
\right).
\label{time-derivative-density}
\end{eqnarray}
In some cases, a set of $\eta_i^j$'s may be obtained by solving Eq.~(\ref{orb_eqs_P}). In the MCTDHF theory, however, it is well known that Eq.~(\ref{orb_eqs_P}) does not need to be solved, or indeed can not, and the $\eta_i^j$'s are thus often chosen to be zero \cite{Beck2000, Meyer2010}. Such freedom does not exist in the TD-RASSCF theory because the Fock space $\mathcal{V}_{\rm RAS}$ does not consist of all possible configurations. It is always possible, however, to set $\eta_{j}^{i}=0$ within each subspace $\mathcal{P}_K$ ($K=0,1$, and $2$), i.e., when two orbitals $\phi_i(t)$ and $\phi_j(t)$ belong to the same subspace (see Fig.~\ref{fig_orbital_ras}). In the TD-RASSCF theory, generally Eq.~(\ref{orb_eqs_P}) needs to be solved for combinations $(i',j'')$ to determine $\eta^{j''}_{i'}$ $\big(=-{(\eta_{j''}^{i'})}^*\big)$, where indices with single and double prime hereafter mean that the orbitals labeled by them belong to different subspaces. Substituting Eqs.~(\ref{Hamiltonian1}) and (\ref{definition_D}) into Eq.~(\ref{orb_eqs_P}) and computing commutators, the explicit expression of the $\mathcal{P}$-space orbital equations reads
\begin{eqnarray}
\sum_{k''l'} (h^{k''}_{l'}-i\eta^{k''}_{l'}) A^{l'j''}_{k''i'}
+\sum_{klm}
(v^{j''m}_{kl} \rho^{kl}_{i'm} -v^{kl}_{i'm} \rho^{j''m}_{kl})
=i\dot{\rho}_{i'}^{j''},
\nonumber \\
\label{P_orb_eqs5}
\end{eqnarray}
where
\begin{eqnarray}
A^{l'j''}_{k''i'}
=
\langle\Psi|\big[c_{i'}^{\dagger}c_{j''},c_{k''}^{\dagger}c_{l'}\big]|\Psi\rangle
=
\delta^{j''}_{k''}\rho^{l'}_{i'}-\delta^{l'}_{i'}\rho^{j''}_{k''}.
\end{eqnarray}
To carry out time propagation of the wave packet, the set of equations of motion, i.e., Eqs.~(\ref{amp_orb}), (\ref{Q_orbital_eq1}), and (\ref{P_orb_eqs5}), need to be solved. Notice that the numerical integration of the amplitude and the $\mathcal{P}$-space orbital equations requires elaborate implicit integration schemes: for solving the amplitude equations (\ref{amp_orb}) to compute the values of $\{\dot{C}_I\}_{I\in\mathcal{V}_{\rm RAS}}$, the values need to be known beforehand to prepare the values of $\eta_{i'}^{j''}$, because $\eta_{i'}^{j''}$ is considered to be a function of $\{\dot{C}_I\}_{I\in\mathcal{V}_{\rm RAS}}$ via $\dot{\rho}_{i'}^{j''}$. One way to easily circumvent the use of implicit integrations is by taking only into consideration even occupation numbers in the $\mathcal{P}_2$-space, which removes $\dot{\rho}_{i'}^{j''}$, and thus the $\mathcal{P}$-space orbital equations result in
\begin{eqnarray}
\sum_{k''l'} (h^{k''}_{l'}-i\eta^{k''}_{l'}) A^{l'j''}_{k''i'}
+\sum_{klm}
(v^{j''m}_{kl} \rho^{kl}_{i'm} -v^{kl}_{i'm} \rho^{j''m}_{kl})
=0.
\nonumber \\
\label{P_orb_eqs6}
\end{eqnarray}
The amplitude and the $\mathcal{P}$-space orbital equations are now separable and can be easily solved by usual explicit integration algorithms. In this work, all the calculations are based on Eq.~(\ref{P_orb_eqs6}), by which all the singly-excited configurations are abandoned. However, it should be noted that even within this scheme, the wave packet partly includes single electron excitation processes due to the time-dependent Brillouin's theorem [Eqs.~(\ref{Q_orb_eqs0}) and (\ref{orb_eqs_P})]. In other words, by solving the $\mathcal{Q}$- and $\mathcal{P}$-space orbital equations for a given set of the $\mathcal{P}$-space orbitals $\{\phi_i(t)\}_{i=1}^{2M}$, we obtain a new set of orbitals $\{\phi_i(t+dt)\}_{i=1}^{2M}$, which are variationally optimized to take into account the single electron processes between the $\mathcal{P}$- and $\mathcal{Q}$-spaces and among the $\mathcal{P}_0$-, $\mathcal{P}_1$-, and $\mathcal{P}_2$-spaces at any instant of time $t$.



\section{\label{Pspace} Parametric redundancy}

\subsection{$\bm{\mathcal{P}}$-space orbital equations revisited}

In the preceding section, the derivation of the $\mathcal{P}$-space orbital equations was briefly sketched followed by a discussion of how to solve them. We now revisit certain details of the derivation. Consider the substitution of Eq.~(\ref{time-derivative-density}) into Eq.~(\ref{orb_eqs_P}): by using a formal expression of the amplitude equations (\ref{amp_eqs1}), we arrive at another expression of the $\mathcal{P}$-space orbital equations
\begin{eqnarray}
\langle\Psi|(iD-H)(1-\Pi)|\Psi^{i}_{j}\rangle
-
\langle\Psi_{i}^{j}|(1-\Pi)(iD-H)|\Psi\rangle
=0,
\nonumber \\
\label{P_orb_eqs03}
\end{eqnarray}
with a projection operator defined by
\begin{eqnarray}
\Pi=
\sum_{I\in\mathcal{V}_{\rm RAS}}|\Phi_I\rangle\langle\Phi_I|.
\label{Pi}
\end{eqnarray}
In the MCTDHF theory, i.e., when $\mathcal{V}_{\rm RAS}$ is replaced by $\mathcal{V}_{\rm FCI}$ defined above Eq.~(\ref{total_wave_CI}), since the Fock space $\mathcal{V}_{\rm FCI}$ includes all possible configurations, the left hand side of the expression~(\ref{P_orb_eqs03}) is zero because  $(1-\Pi)|\Psi^{i}_{j}\rangle=\langle\Psi_{i}^{j}|(1-\Pi)=0$ for any combination of $i$ and $j$. Hence, the $\mathcal{P}$-space orbitals are completely undetermined, and one can therefore choose arbitrary anti-hermitian matrices for $\eta_{i}^{j}$ (see, e.g., Refs.~\cite{Beck2000,Meyer2010}). This fact stems from the non-uniqueness of $\{C_I\}_{I\in\mathcal{V}_{\rm FCI}}$ and $\{\phi_i\}_{i=1}^{2M}$. As is well known in the time-independent SCF theory, a unitary transformation of the orbitals 
\begin{eqnarray}
|\phi_i\rangle 
= \sum_j |\phi_j'\rangle G_{ji}
\label{Orb_G}
\end{eqnarray}
together with the transformation of the expansion coefficients
\begin{eqnarray}
C_I 
=
\sum_{j_1}\cdots\sum_{j_{N_{\rm e}}} G^{-1}_{i_1j_1}\cdots G^{-1}_{i_{N_{\rm e}}j_{N_{\rm e}}}
C_J'
\label{Amp_G}
\end{eqnarray}
keeps the wave function invariant. This property at a certain moment of time is called parametric redundancy in time-independent quantum chemistry \cite{Helgaker2000}. In the time-dependent formulation, however, it is of importance as well to consider the time evolution of the unitary transformation:
\begin{eqnarray}
\eta^j_i
&=&
\langle\phi_j|\dot{\phi}_i\rangle
\nonumber \\
&=&
\sum_{kl}
G^{-1}_{lj}\dot{G}_{ki}
\langle\phi'_{l}|\phi'_{k}\rangle
+
\sum_{kl}
G^{-1}_{lj}G_{ki}
\langle\phi'_{l}|\dot{\phi}'_{k}\rangle
\nonumber \\
&=&
\sum_{k}
G^{-1}_{kj}\dot{G}_{ki}
+
\sum_{kl}
G^{-1}_{lj}G_{ki}
{\eta'}^{l}_{k},
\end{eqnarray}
which is formally solved in matrix form
\begin{eqnarray}
{\bm G}(t)=\mathscr{T}\exp\left[\int_0^t \Big({\bm \eta}(t')-{\bm \eta}'(t')\Big)dt' + {\bm \Delta} \right],
\label{G(t)}
\end{eqnarray}
where ${\bm \Delta}$ is a constant anti-hermitian matrix and $\mathscr{T}$ denotes time-ordering. This equation means that between any two sets of orbitals, there exists a unitary transformation at any moment of time and therefore it ensures a unique description of the time-dependent dynamics by using an arbitrary set of orbitals. This kind of geometrical structure is an advanced concept of the parametric redundancy and is known as principal bundle, in which the gauge map defined by ${\bm \eta}(t)$ exists \cite{Lubich2008,Kvaal2011}. Exploiting the gauge degree of freedom, usually the gauge is fixed such that $\eta^j_i=0$ at all times to make the $\mathcal{P}$-space orbital equations vanish and simplify the system of differential equations. Another useful gauge choice is employing $\eta^j_i=-ih^j_i$ by which the use of a larger time step is sometimes allowed in time propagation \cite{Caillat2005, Beck2000, Meyer2010}. 

In the TD-RASSCF theory, the expression (\ref{P_orb_eqs03}) is still a trivial identity for combinations $(i,j)$ belonging to the same subspace in $\mathcal{P}$, i.e., if $\phi_i(t)\in\mathcal{P}_K$ and $\phi_j(t)\in\mathcal{P}_K$ ($K=0,1$, and $2$). The time-dependent orbital rotations within each subspace are hence undetermined. However, this is not the case for the combinations $(i',j'')$, i.e., if $\phi_{i'}(t)\in\mathcal{P}_{K}$ and $\phi_{j''}(t)\not\in\mathcal{P}_{K}$. This is because the unitary transformations (\ref{Orb_G}) and (\ref{Amp_G}) can now be carried out only within each subspace, and Eq.~(\ref{G(t)}) is still true in each subspace. Therefore, fixing the gauge such that $\eta^{j}_{i}=0$ is satisfied, what remains is to determine the off-diagonal block elements $\eta^{j''}_{i'}$ $\big(=-{(\eta_{j''}^{i'})}^*\big)$ by solving Eq.~(\ref{P_orb_eqs5}). Here it is important to notice that $\rho^{j''}_{i'}=\dot{\rho}^{j''}_{i'}=0$ when either $i'$ or $j''$ denotes the index of a $\mathcal{P}_0$-space orbital. Furthermore, by taking only into consideration even occupation numbers in the $\mathcal{P}_2$-space, $\rho^{j''}_{i'}$ and $\dot{\rho}^{j''}_{i'}$ vanish, and the $\mathcal{P}$-space orbital equations thereby result in Eq.~(\ref{P_orb_eqs6}). The amplitude and the $\mathcal{P}$-space orbital equations are now separable. Another prescription to set $\rho^{j''}_{i'}=\dot{\rho}^{j''}_{i'}=0$ is forbiding electron transitions between $\mathcal{P}_1$- and $\mathcal{P}_2$-spaces. This complete-active-space (CAS) scheme \cite{Helgaker2000,Meier1989} gives us formally the same $\mathcal{P}$-space orbital equations \cite{Sato2013}.

\subsection{Related works}

It is informative to briefly discuss two related works. The first one is the MCTDH method with selected configurations (S-MCTDH) \cite{Worth2000}. To simplify the problem, the S-MCTDH method ignores the treatment of the $\mathcal{P}$-space orbital equations, which are thus assumed to be always satisfied, i.e., supposed to be identities, not equations. Although the S-MCTDH method works efficiently for computing absorption spectra of a pyrazine molecule, the method exhibits numerical instability as well for some configuration selections conceivably due to the discard of the $\mathcal{P}$-space orbital equations. The other related method is based on the MCTDHF theory with a truncation of the expansion \cite{Miranda2011}. To reduce the numerical cost, the method employs the time-independent expansion coefficients, i.e., fixed values throughout the time propagation. These two works abandon either the amplitude or the $\mathcal{P}$-space orbital equations as an additional approximation, which lowers the accuracy of the description of the dynamics. Both the amplitude and the $\mathcal{P}$-space orbital equations are exactly treated in the present TD-RASSCF theory, in which, the $\mathcal{P}$-space orbital equations are simple to solve and the computational cost is largely reduced by the RAS scheme. Before closing this section, we emphasize that the gauge degree of freedom due to the principal bundle structure is a key concept behind the treatment of the $\mathcal{P}$-space orbital equations. One can find a discussion of this issue in terms of the OATDCC theory in the supplementary material of Ref.~\cite{Kvaal2012}. 


\section{\label{Numerical tests} Numerical application}

\subsection{\label{Numerical tests of MCTDHF-n} Ground state wave function}

\begin{figure}
\begin{center}
\begin{tabular}{c}
\resizebox{80mm}{!}{\includegraphics{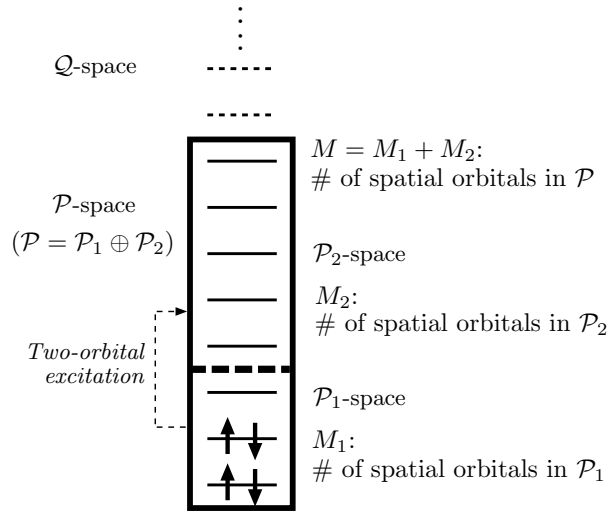}}
\end{tabular}
\caption{
\label{fig2_orbital_ras}
Illustration of the single-particle Hilbert space used in the calculations for the 1D beryllium atom. In this case, the $\mathcal{P}$-space is simply decomposed into two active spaces: only two-orbital transitions between the $\mathcal{P}_1$- and $\mathcal{P}_2$-spaces are now permitted. The numbers of spatial orbitals in the $\mathcal{P}_1$- and $\mathcal{P}_2$-spaces are expressed by $M_1(\ge 1)$ and $M_2(\ge 0)$, respectively, and the total number by $M$ $(=M_1+M_2)$. In this illustrative example of $(M,M_1)=(8,3)$, the electrons are in the lowest energy configuration in the $\mathcal{P}_1$-space. Notice the special case $M_1=M$, where the $\mathcal{P}_2$-space disappears and the present RAS scheme thereby becomes the MCTDHF theory. In the application to the 1D helium atom, the same partitioning in the $\mathcal{P}$-space was used.
}
\end{center}
\end{figure}

We investigate $N_{\rm e}$-electron atoms to demonstrate the computational efficiency and analyze the convergence property of the TD-RASSCF theory by proof-of-principle calculations. The atoms are modeled by 1D systems with soft-core Coulomb potential: The one-body operator in Eq.~(\ref{one-body1}) is
\begin{eqnarray}
h(x)=-\frac{1}{2}\frac{d^2}{dx^2}+V(x)
\label{one-body}
\end{eqnarray}
with
\begin{eqnarray}
V(x)=-\frac{Z}{\sqrt{x^2+1}},
\end{eqnarray}
where $Z=N_{\rm e}=2$ for describing a helium atom \cite{Hochstuhl2010,Balzer2010,Balzer2010b} and $Z=N_{\rm e}=4$ for a beryllium atom \cite{Bonitz2010,Hochstuhl2010b}. The two-body operator in Eq.~(\ref{two-body1}) is
\begin{eqnarray}
v(x_1,x_2)=\frac{1}{\sqrt{(x_1-x_2)^2+1}}.
\end{eqnarray}
In this section, the RAS scheme is simplified by eliminating the inactive-core space $\mathcal{P}_0$ as shown in Fig.~\ref{fig2_orbital_ras}: only two-electron transitions are allowed between the $\mathcal{P}_1$- and $\mathcal{P}_2$-spaces, in which the numbers of spatial orbitals are $M_1(\ge1)$ and $M_2(\ge0)$, respectively, and the total number is $M(=M_1+M_2)$. In this scheme, the TD-RASSCF theory is equivalent to the MCTDHF theory when $M_1=M$.

\begin{table*}
\caption{\label{table_Be}
Ground state energy in atomic units of the 1D beryllium atom ($Z=N_{\rm e}=4$) for different combinations of $(M,M_1)$ (see caption of Fig.~\ref{fig2_orbital_ras}). The integers in parentheses below each energy value show the number of configurations used in the calculation. The non-monotonic improvement of the energy for fixed $M$ and increasing $M_1$ is discussed in the text.
}
\begin{ruledtabular}
\begin{tabular}{rccccccc}
$M$\hspace{0.0mm}		
		& 2			& 3			& 4			& 6			& 8			&10 			& 12  		\\ \hline \\
$M_1=1$	&			& $-6.771296$	& $-6.775354$	& $-6.776631$	& $-6.776764$	& $-6.776780$	& $-6.776782$  \\
		&			& $(5)$		& $(18)$		& $(125)$		& $(490)$		& $(1377)$	& $(3146)$	\\
$2$	& $-6.739450$	& $-6.771296$	& $-6.779805$	& $-6.784224$	& $-6.784501$	& $-6.784529$	& $-6.784533$  \\
		& $(1)$		& $(5)$		& $(19)$		& $(77)$		& $(175)$		& $(313)$		& $(491)$ 	 	\\
$3$	& 			& $-6.771296$	& $-6.775314$	& $-6.779301$	& $-6.779648$	& $-6.779683$	& $-6.779687$  \\
		& 			& $(9)$		& $(18)$		& $(108)$		& $(294)$		& $(576)$¤	& $(954)$ 	 	\\
$4$	& 		 	&			& $-6.780026$	& $-6.781293$	& $-6.781591$	& $-6.781627$	& $-6.781633$  \\
		& 		 	& 			& $(36)$		& $(112)$		& $(364)$		& $(792)$¤	& $(1396)$ 	 \\
$6$	& 		 	& 			& 			& $-6.784736$	& $-6.784814$	& $-6.784838$	& $-6.784843$  \\
		& 		 	& 			& 			& $(225)$		& $(399)$		& $(981)$¤	& $(1971)$ 	 \\
$8$	& 		 	& 			& 			& 			& $-6.785041$	& $-6.785049$	& $-6.785050$  \\
		& 		 	& 			& 			& 			& $(784)$		& $(1096)$	& $(2144)$ 	 \\
$10$& 		 	&			& 			& 			& 			& $-6.785072$	& $-6.785074$  \\
		& 		 	&			& 			& 			& 			& $(2025)$	& $(2515)$ 	 \\
$12$& 		 	&			& 			& 			& 			&			& $-6.785077$  \\
		& 		 	&			& 			& 			& 			&			& $(4356)$ 	 \\
\end{tabular}
\end{ruledtabular}
\end{table*}

\begin{figure*}
\begin{center}
\begin{tabular}{c}
\resizebox{140mm}{!}{\includegraphics{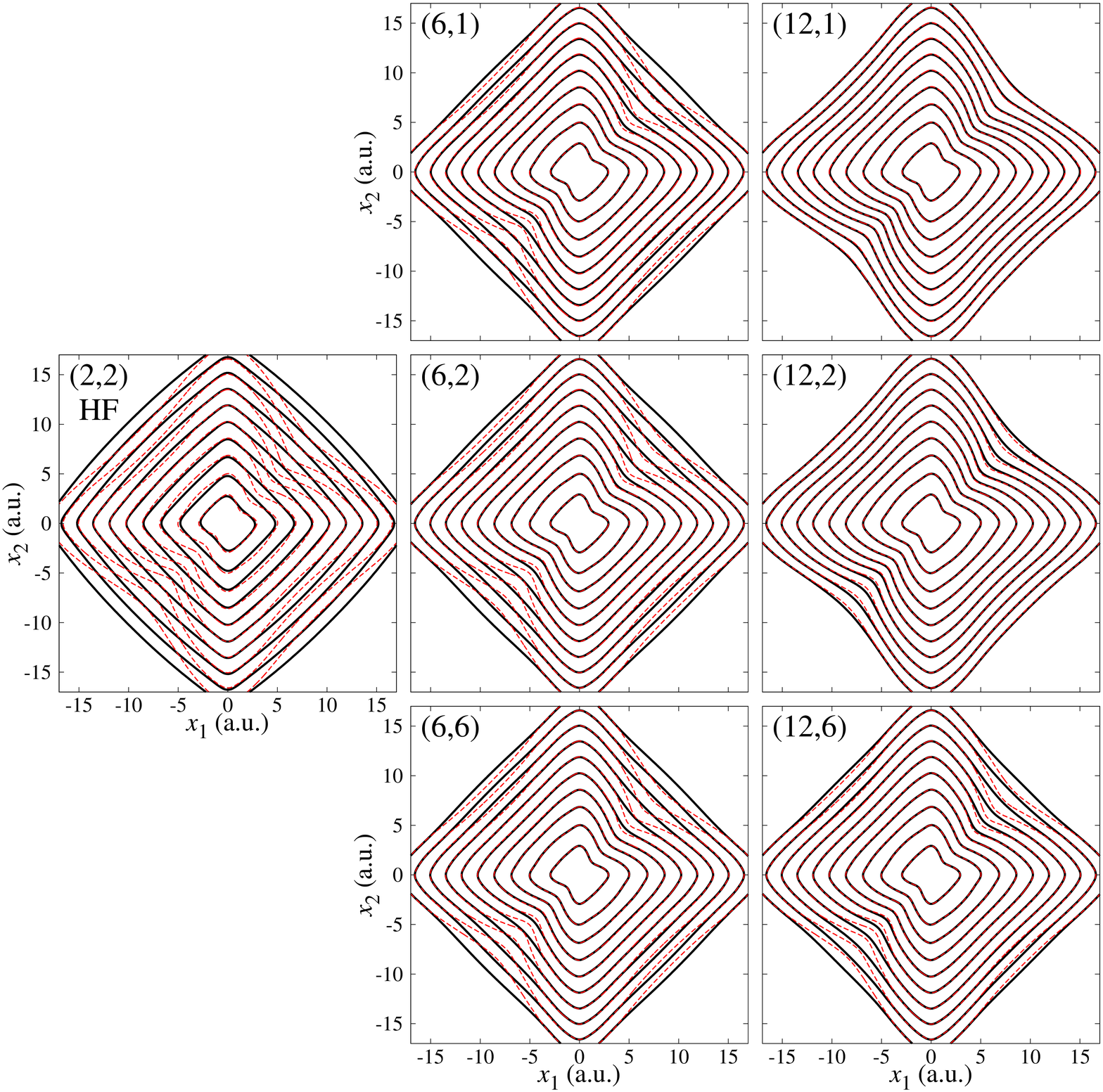}}
\end{tabular}
\caption{
\label{fig_density_Be}
(Color online) Logarithmic contour plot of the spin-averaged two-electron density $\rho(x_1,x_2)$ [Eq.~(\ref{2density})] of the 1D beryllium atom in the ground state for different combinations of $(M,M_1)$ as indicated on each panel. The leftmost result of $(M,M_1)=(2,2)$ corresponds to the HF result. For comparison, all panels include the same dashed (red) lines representing the density calculated with $(M,M_1)=(12,12)$. Contours differ by a factor of 10, indicating $0.1$ for the line of the innermost island. 
}
\end{center}
\end{figure*}

Table \ref{table_Be} lists the ground state energy of the 1D beryllium atom for different combinations of $(M,M_1)$. The results calculated by the MCTDHF method ($M_1=M$) exactly agree with those given in Ref.~\cite{Bonitz2010}. All the results are obtained from imaginary time relaxation in a box $[-25,25]$ discretized by the discrete-variable-representation (DVR) \cite{Light1985} quadrature points, $N_{\rm DVR}=256$, associated with Fourier basis functions. The integer in the parenthesis below each energy value gives the number of configurations. Figure~\ref{fig_density_Be} depicts some selected results for the spin-averaged two-electron density
\begin{eqnarray}
\rho_2(x_1,x_2)
&\equiv&
\frac{1}{4}
\sum_{ijkl}\rho^{jl}_{ik}
\int \!\!d\sigma_1 \!\int \!\!d\sigma_2
\nonumber \\
&& \times
\big\|\phi^{\dagger}_i(z_1) \phi^{\dagger}_k(z_2)\big\|
\big\|\phi_j(z_1) \phi_l(z_2)\big\|,
\label{2density}
\end{eqnarray}
where $\|\cdots\|$ means the normalized Slater determinant (for two-electron systems, $\rho_2(x_1,x_2)$ is equivalent to the absolute square of the spatial wave function). 

\begin{figure*}
\begin{center}
\begin{tabular}{c}
\resizebox{115mm}{!}{\includegraphics{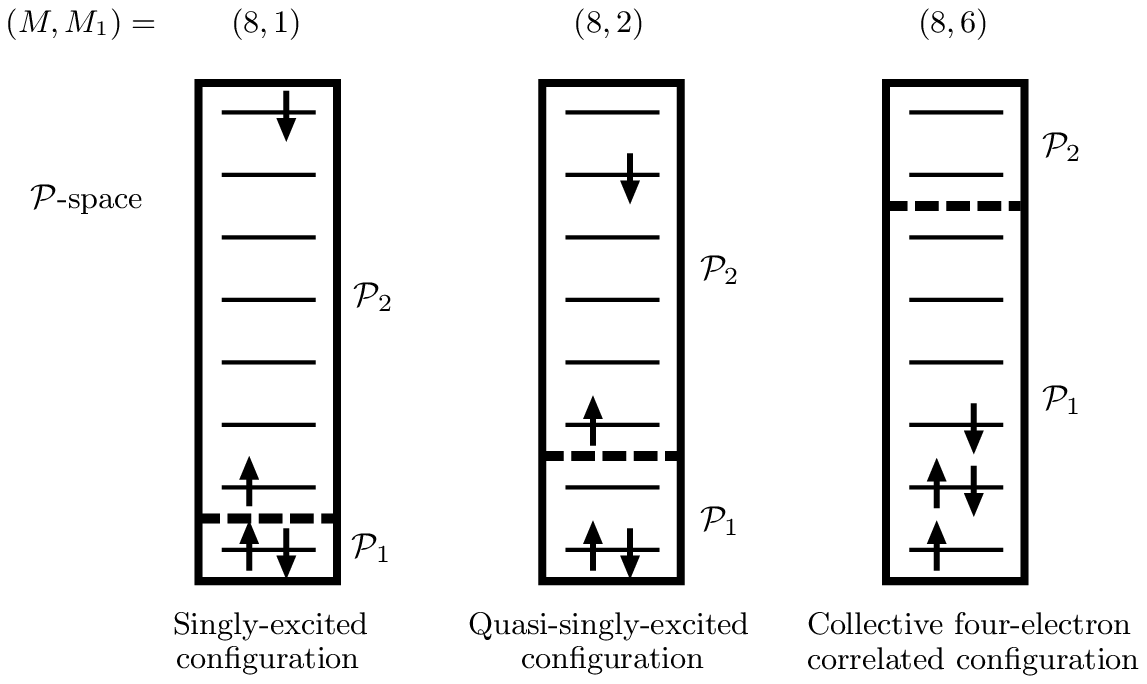}}
\end{tabular}
\caption{
\label{fig3_orbital_ras}
Concept of the excited configurations in the TD-RASSCF theory for the 1D beryllium atom (see also Fig.~\ref{fig_density_Be}). These are typical excited configurations for $(M,M_1)=(8,1)$, $(8,2)$, and $(8,6)$ from left to right, respectively. Relative to the lowest energy configuration, the singly-excited configurations are realized only for $(M,M_1)=(8,1)$. The calculation of $(M,M_1)=(8,2)$ includes the doubly-excited configurations called quasi-singly-excited configurations, where, in the $\mathcal{P}_2$-space, one of the two excited electrons remains in a low-energy orbital near the nucleus but the other occupies a high-energy orbital far away from the nucleus. These configurations are excluded from the method of $(M,M_1)=(8,6)$, which, however, includes collective four-electron correlated configurations near the nucleus. 
}
\end{center}
\end{figure*}

In Table \ref{table_Be} and Fig.~\ref{fig_density_Be}, for a fixed $M_1$, the larger $M$, i.e., the more active orbitals, variationally the more accurate a result is obtained. On the other hand, for a fixed $M$, the use of larger $M_1$ does not necessarily gives more accurate results, because the exclusion of single-orbital excitations from $\mathcal{P}_1$- to
$\mathcal{P}_2$-space makes the theory non-variational with respect to the position of the partition.
To clarify the physics behind this convergence behavior, consider how the RAS scheme is working: the four electrons are firstly distributed in all possible manners in the $\mathcal{P}_1$-space, from which all possible two-orbital transitions to the $\mathcal{P}_2$-space take place, as shown in Fig.~\ref{fig3_orbital_ras}, where typical configurations realized for $(M,M_1)=(8,1)$, $(8,2)$, and $(8,6)$ are illustrated. Notice that, relative to the lowest energy configuration, the singly-excited configurations are realized only for $(M,M_1)=(8,1)$. The wave function of $(M,M_1)=(8,2)$, however, includes doubly-excited configurations hereafter called \textit{quasi-singly-excited} configurations, in which one of the two excited electrons remains in a low-energy orbital near the nucleus but the other occupies a high-energy orbital far away from the nucleus. These singly- and quasi-singly-excited configurations are excluded from the method of $(M,M_1)=(8,6)$, which has an advantage to take into account \textit{collective four-electron correlated} configurations near the nucleus. In short, the larger $M_1$, i.e., the more upward the partition shifts, the more configurations in the $\mathcal{P}_1$-space but the less configurations in the $\mathcal{P}_2$-space the wave function includes.

Look at the line of $M_1=2$ in Table \ref{table_Be}, where the number of configurations is largely reduced, and thereby making the computation time shorter. In this line a slow convergence with respect to $M$ can be seen: starting from the Hartree-Fock (HF) energy, the energy value decreases and eventually becomes $-6.784533$ at $M_1=12$, which still differs from the value $-6.785077$ obtained from the $(M,M_1)=(12,12)$ calculation. The slow convergence property is more pronounced in the calculations using $M_1=1$. This is due to the lack of the collective four-electron correlated configurations for describing the complex electron correlation near the nucleus. On the other hand, for a fixed $M(\ge 6)$, an energy value calculated with $M_1=6$ is always more accurate than corresponding ones with $M_1=1$ and $2$. However, against the superiority of the use of $M_1=6$ to $M_1=1$ and $2$ for calculating the ground state energy, Fig.~\ref{fig_density_Be} indicates an opposite view; the uses of $M_1=1$ and $2$ are seemingly superior to $M_1=6$ for a more accurate description of the two-electron density in logarithmic scale. In a region far from the nucleus and especially around $x_1\simeq x_2$, the density is remarkably well described in the calculations with $M_1=1$ and $2$. This observation so far interestingly indicates that taking into consideration  the collective four-electron correlated configurations is crucial for obtaining an accurate ground state energy, while the singly- or quasi-singly-excited configurations are important for the details of the  electron density in the region far from the nucleus, which is  where the tunneling ionization by strong lasers takes place, as will be discussed in the next subsection. 

Finally we consider an application to the 1D helium atom ($Z=N_{\rm e}=2$). Under the same numerical condition, a direct solution of the Schr\"odinger equation provides the exact ground state energy $-2.238257$. On the other hand, rapid convergence of the ground state energy is observed in the MCTDHF calculation with increasing $M$; starting from the HF energy $-2.224210$, almost converged energy is already obtained with $M=7$ (see also Ref.~\cite{Hochstuhl2010}, which lists the ground state energy of the same 1D helium atom for several values of $M$). Here it should be noted that, for two-electron systems, the wave function in the present TD-RASSCF scheme is invariant with respect to the position of the partition (see Appendix~\ref{Remove}). This is also the case for general $N_{\rm e}$-electron atoms when $M=N_{\rm e}/2+1$, because there are two holes which play the same role as the two electrons in two-electron systems. In the case of the 1D beryllium atom with $M=3$, the TD-RASSCF calculations using $M_1=1$, $2$, and $3$ hence provide equal  ground state energy as shown in Table~\ref{table_Be}. 

\subsection{\label{HHG} High-order harmonic generation}

To investigate the performance of the TD-RASSCF theory in a truely time-dependent setting, we consider the dynamics of the 1D beryllium model atom ($Z=N_{\rm e}=4$) interacting with a few-cycle near-infrared  laser field. The effect of the laser is described by adding the dipole interaction in the length gauge to the one-body Hamiltonian (\ref{one-body}) as
\begin{eqnarray}
h(x,t)=-\frac{1}{2}\frac{d^2}{dx^2}+V(x)+xF(t)-iW(x),
\label{h_laser}
\end{eqnarray}
with the laser field expressed by
\begin{eqnarray}
F(t)=F_0\cos^2\left(\frac{\pi t}{T}\right)\cos \omega t,
\hspace{2mm} (-T/2\leq t\leq T/2).
\end{eqnarray}
Here $F_0$ is the electric field strength, $\omega$ the angular frequency. All the calculations in this section were carried out using a larger box $[-300,300(\equiv L)]$ discretized by $N_{\rm DVR}=2048$ points. The real time propagation was implemented by the fourth-order Runge-Kutta method with time step $\Delta t = 0.005$. To cure the electron reflections at the edges of the box, Eq.~(\ref{h_laser}) includes the complex absorbing potential (CAP) function defined by $W(x)=1-\cos\Big\{\pi(|x|-x_{\rm cap})/\big[2(L-x_{\rm cap})\big]\Big\}$ with $x_{\rm cap}=250$ for $|x|>x_{\rm cap}$ and zero otherwise \cite{Zanghellini2006}. To keep the calculations stable, a further numerical technique is needed. In the $\mathcal{Q}$-space orbital equations (\ref{Q_orbital_eq1}) the density matrix is regularized by the following substitution
to prevent it from being singular
\begin{eqnarray}
{\bm \rho}_{\rm reg}\equiv
{\bm \rho}
+\epsilon \exp\big(-{\bm \rho}/\epsilon\big)
\end{eqnarray}
with a small constants $\epsilon=10^{-10}$ \cite{Beck2000}. The same regularization method was used for the tensor $A^{l'j''}_{k''i'}$ in the $\mathcal{P}$-space orbital equations (\ref{P_orb_eqs6}). The validity of the numerical data shown below was checked by carrying out the same calculations with using larger boxes, denser DVR quadrature points, smaller time steps, and different values of the CAP parameter, $x_{\rm cap}$.

As an important observable, the HHG spectra are calculated from the dipole moment in the acceleration form
\begin{eqnarray}
S(\Omega)=
\Bigg|
\int^{T/2}_{-T/2}
\langle\Psi(t)|
\sum_{\kappa=1}^{N_{\rm e}}
\left(
-\frac{d}{dx_{\kappa}}V(x_{\kappa})
\right)
|\Psi(t)\rangle e^{i\Omega t}
dt
\Bigg|^2,
\nonumber \\
\label{S-ac}
\end{eqnarray}
which is supposed to be more favorable than in the length form especially when a CAP function is used \cite{Sukiasyan2009,Sukiasyan2010}. There are also other superiorities for the use of the acceleration form to the length form as discussed in Ref. \cite{Gordon2006}. Note that, in Eq.~(\ref{S-ac}), the laser electric field is excluded since it does not contribute to the HHG spectrum.

\begin{figure*}
\begin{center}
\begin{tabular}{cc}
\resizebox{170mm}{!}{\includegraphics{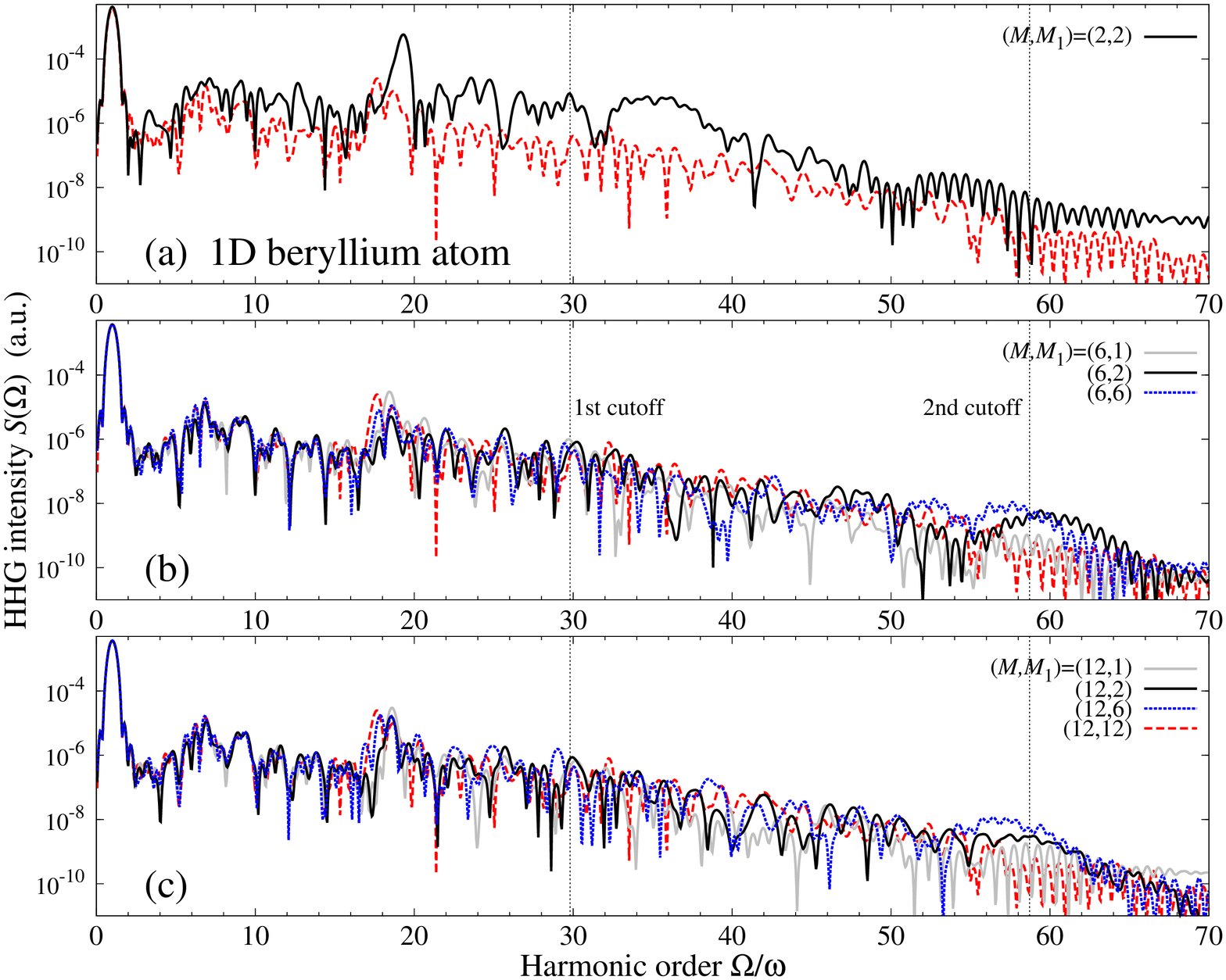}}
\end{tabular}
\caption{
\label{fig_Be_HHG}
(Color online) HHG spectra of the 1D beryllium atom calculated for different combinations of $(M,M_1)$ (see caption of Fig.~\ref{fig2_orbital_ras}). The result of $(M,M_1)=(2,2)$ shown by the solid black line in (a) corresponds to the TDHF result. For comparison, all panels include the same dashed (red) line corresponding to the result of $(M,M_1)=(12,12)$. The laser pulse is specified by the parameters: $F_0=0.0755$ ($2.0\times10^{14}$ Wcm$^{-2}$), $\omega=0.0570$ ($800$ nm), and $T=331$ ($3$ cycles). The first and second cutoff energies are estimated to be $29.8\omega$ and $58.7\omega$, respectively, as shown by vertical dotted lines (see text).
}
\end{center}
\end{figure*}

Figure~\ref{fig_Be_HHG} displays the HHG spectra of the 1D beryllium atom induced by a laser pulse specified by $F_0=0.0755$ ($2.0\times10^{14}$ Wcm$^{-2}$), $\omega=0.0570$ ($800$ nm), and $T=331$ ($3$ cycles). For comparison, all panels include the same dashed (red) line representing the MCTDHF result calculated with $(M,M_1)=(12,12)$, which 
includes most electron correlation. In all cases a double plateau structure appears due to the many-electron effect and the use of the short pulse. The vertical dotted lines in Fig.~\ref{fig_Be_HHG} indicate the positions of the first and second cutoffs estimated based on the three-step model by solving the classical equations of motion for a free electron in the laser field as follows: Within the second laser cycle, a liberated electron returns back to the parent ion with the maximum kinetic energy $3.15U_{\rm p}$, which accounts for the first cutoff as $3.15U_{\rm p}+I_{\rm p}^{(1)}=29.8\omega$. Here the first ionization potential is estimated from the highest occupied orbital energy $-0.3127982(\equiv -I^{(1)}_{\rm p})$ in the HF approximation, and $U_{\rm p}=F_0^2/(4\omega^2)=0.439$ is the ponderomotive potential. Slightly after this moment of time, another electron already ejected in the first laser cycle returns to the parent ion. Treating the two electrons coherently with neglecting the electron repulsion \cite{Koval2007}, the nonsequential double recombination emits a photon having the maximum energy $5.04U_{\rm p}+I_{\rm p}^{(1)}+I_{\rm p}^{(2)}=58.7\omega$, which could roughly explain the second cutoff. Here the sum of the first and second ionization potentials is estimated to be $I_{\rm p}^{(1)}+I_{\rm p}^{(2)}=E^{2+}_{\rm g}-E_{\rm g}=1.129997$, where $E_{\rm g}$ and $E^{2+}_{\rm g}$ are the ground state energies of the 1D beryllium atom and its dication Be$^{2+}$, respectively, obtained by the HF calculations. 

We start the discussion of the spectra in Fig.~\ref{fig_Be_HHG} by investigating the structure of the first plateau ($0<\Omega/\omega<30$). Although the overall shape of the HHG spectra is similar in all calculations, the TDHF result, i.e., the result of $(M,M_1)=(2,2)$ in Fig.~\ref{fig_Be_HHG} (a) shows a significant disagreement with the result of $(M,M_1)=(12,12)$. This failure is because the creation of the first plateau is mainly due to the single ionization and recombination processes, which are not included explicitly in the TDHF wave function. The increase of $M$ removes this shortcoming, and the variational improvement of the accuracy is thus apparent in Figs.~\ref{fig_Be_HHG} (b) and (c), in which all the results are in reasonable agreement. Here the agreement especially among the $(M,M_1)=(6,1)$, $(6,2)$, $(12,1)$, $(12,2)$, and $(12,12)$ results importantly indicates that the singly- and quasi-singly-excited configurations play a leading role to reproduce the first plateau. Around $15<\Omega/\omega<30$, however, all the results show small disagreements, which require further analysis beyond the scope of the present work.

Next look into the second plateau ($30<\Omega/\omega<60$). The TDHF result in Fig.~\ref{fig_Be_HHG} (a) differs much from the result of $(M,M_1)=(12,12)$. As discussed so far, the nonsequential double recombination roughly estimates the creation of the second plateau. Although the TDHF wave function takes into account the double continua, the two liberated electrons are always in the same spatial orbital, which results in the poor accuracy of the TDHF method. On the other hand, in Fig.~\ref{fig_Be_HHG} (b) and (c), the second plateau is roughly reproduced by all methods, including the $(M,M_1)=(6,2)$ and $(12,2)$ methods despite of the large reduction of the number of configurations in these approach. It seems to be a formidable task to exactly reproduce the fine structure. The rich structure in the second plateau especially around the second cutoff is due to the interference among several quantum trajectories of the two electrons coming back to the parent ion \cite{Koval2007}. It is thus still questionable whether the convergence is completely achieved even in the calculation with $(M,M_1)=(12,12)$. Notice that, in Ref.~\cite{Jordan2008}, HHG spectra were calculated for a four-electron molecule using a similar laser pulse by the MCTDHF method. In this related work, however, all the results including the TDHF one show reasonable agreement in both the first and second plateaus. The convergence behavior of the MCTDHF calculations will thus sensitively depend on the system, as well as the parameters of the driving laser. 
However, as shown in Fig.~\ref{fig_Be_HHG} (c), a good agreement between the MCTDHF and the TD-RASSCF results is observed in the first plateau despite the fact that largely different sets of electronic configurations are used in the two methods. This is an indicator of the convergence of each calculation for describing the single-electron continuum states. Furthermore, it is safe to say that the reasonable agreement in the second plateau indicates some account of the double continuum.

\section{\label{Conclusion}Conclusions and outlooks}

In this paper, a new theoretical framework for describing time-dependent many-electron dynamics, denoted TD-RASSCF theory, was proposed and formulated on the basis of the time-dependent variational principle. The key concepts of the theory are the use of the time-dependent orbitals and the truncation of the CI-expansion by incorporating the RAS scheme. Abandoning the full-CI expansion gives rise to important changes in the formulation as compared to the MCTDHF theory. The TD-RASSCF theory thereby requires us to solve the $\mathcal{P}$-space orbital equations. To make the amplitude and the $\mathcal{P}$-space orbital equations separable, we only allow transitions of even numbers of orbitals between the $\mathcal{P}_1$- and $\mathcal{P}_2$-spaces. In a proof-of-principle application to the 1D beryllium atom, the TD-RASSCF method exhibited a reasonable convergence behavior with accumulating the number of active orbitals in both calculations of the ground state wave function and the HHG spectra induced by an intense laser pulse. By shifting the position of the partition between the two active spaces and changing the number of active orbitals in each subspace, one can flexibly take into account the electron correlation needed for describing the phenomena of interest. This flexibility and the accompanying gain in computational efforts allow us to promote  the TD-RASSCF theory as a promising method for  application in  larger atoms and molecules, beyond the reach of methods based on full CI expansions.

\begin{acknowledgments}
It is a pleasure to thank Dr. Sebastian Bauch (Aarhus University), Dr. Simen Kvaal (University of Oslo), and Dr. Takeshi Sato (the University of Tokyo) for many useful discussions. This work was supported by the Danish Research Council (Grant No. 10-085430) and an ERC-StG (Project No. 277767-TDMET).
\end{acknowledgments}

\appendix


\section{\label{Remove} Wave function of two-electron systems}

Consider a two-electron system in the MCTDHF method with $M(\ge 2)$ active spatial-orbitals. Expressing the index of the active orbitals $|\phi_{i}\rangle$ by $i=1{\uparrow},1{\downarrow},\cdots, M{\uparrow},M{\downarrow}$, the wave function reads
\begin{eqnarray}
|\Psi\rangle
=
|\Phi\rangle
&&+C_{M-1}\Big(|\phi_{1\uparrow}\phi_{M-1\downarrow}\rangle-|\phi_{1\downarrow}\phi_{M-1\uparrow}\rangle\Big)
\nonumber \\
&&
+C_M\Big(|\phi_{1\uparrow}\phi_{M\downarrow}\rangle-|\phi_{1\downarrow}\phi_{M\uparrow}\rangle\Big),
\end{eqnarray}
where $|\Phi\rangle$ means the rest of the Slater determinants. Defining two new orbitals by $|\phi'_{M-1\uparrow(\downarrow)}\rangle\equiv \cos\theta |\phi_{M-1\uparrow(\downarrow)}\rangle-\sin\theta |\phi_{M\uparrow(\downarrow)}\rangle$ and $|\phi'_{M\uparrow(\downarrow)}\rangle\equiv \sin\theta |\phi_{M-1\uparrow(\downarrow)}\rangle+\cos\theta |\phi_{M\uparrow(\downarrow)}\rangle$ such that the condition $C_{M-1}\sin\theta+C_{M}\cos\theta=0$ is fulfilled, for instance, the wave function is rewritten as
\begin{eqnarray}
|\Psi\rangle
=
|\Phi\rangle
+C'_{M-1}\Big(|\phi_{1\uparrow}\phi'_{M-1\downarrow}\rangle-|\phi_{1\downarrow}\phi'_{M-1\uparrow}\rangle\Big),
\end{eqnarray}
where $C'_{M-1}\equiv C_{M-1}\cos\theta-C_{M}\sin\theta$. Proceeding with the orbital manipulations, one can eliminate any or even all of singly-excited configurations relative to the lowest energy configuration. In the TD-RASSCF scheme, all the singly-excited configurations from $\mathcal{P}_1$ to $\mathcal{P}_2$ are likewise removable. Thus the wave function in the TD-RASSCF theory is invariant with respect to the position of the partition between $\mathcal{P}_1$ and $\mathcal{P}_2$. Finally note that, exploiting this flexible structure of the two-electron wave function, one can ultimately arrive at the expression of the wave function in terms of geminals \cite{Kutzelnigg1964} instead of orbitals.











\end{document}